\pdfoutput=1

\documentclass[11pt]{article}

\usepackage[preprint]{acl}

\usepackage{times}
\usepackage{latexsym}

\usepackage[T1]{fontenc}

\usepackage[utf8]{inputenc}

\usepackage{microtype}

\usepackage{inconsolata}

\usepackage{graphicx}
\usepackage{makecell} 
\usepackage{multirow}
\usepackage{booktabs}
\usepackage{xspace}
\usepackage{amssymb}
\usepackage{booktabs}
\usepackage{multirow}
\usepackage{arydshln}
\usepackage{amssymb}
\usepackage{hhline}
\usepackage{tcolorbox}
\usepackage{verbatim}
\usepackage{listings}
\usepackage{fancyvrb}
\usepackage{xcolor}
\usepackage{tikz}
\usepackage{graphicx}
\usepackage{subfigure}
\usepackage{fontawesome}
\newcommand{\codev}{\textsc{CodeV}\xspace}

\title{\codev: Issue Resolving with Visual Data}

\author{Linhao Zhang$^{1}$ \quad Daoguang Zan$^{2}$\thanks{Corresponding authors}  \quad Quanshun Yang$^{1}$ \quad Zhirong Huang$^{2}$ \\ 
\textbf{Dong Chen$^{3}$ \quad Bo Shen$^{3}$ \quad Tianyu Liu$^{4}$ \quad Yongshun Gong$^{1}$ \quad Pengjie Huang$^{5}$} \\
\textbf{Xudong Lu$^{1\ast}$ \quad Guangtai Liang$^{3}$ \quad Lizhen Cui$^{1}$ \quad Qianxiang Wang$^{3}$}\\
$^{1}$Shandong University 
\quad $^{2}$Chinese Academy of Science  \\
\quad $^{3}$Huawei Co., Ltd
\quad $^{4}$Peking University
\quad $^{5}$Lingzhi-zhiguang Co., Ltd 
\\ 
}

\begin{document}
\maketitle
\begin{abstract}
Large Language Models (LLMs) have advanced rapidly in recent years, with their applications in software engineering expanding to more complex repository-level tasks. GitHub issue resolving is a key challenge among these tasks. While recent approaches have made progress on this task, they focus on textual data within issues, neglecting visual data. However, this visual data is crucial for resolving issues as it conveys additional knowledge that text alone cannot. We propose \codev, the first approach to leveraging visual data to enhance the issue-resolving capabilities of LLMs. \codev resolves each issue by following a two-phase process: data processing and patch generation. To evaluate \codev, we construct a benchmark for visual issue resolving, namely Visual SWE-bench. 
Through extensive experiments, we demonstrate the effectiveness of \codev, as well as provide valuable insights into leveraging visual data to resolve GitHub issues\footnote{\url{https://github.com/luolin101/CodeV}}.
\end{abstract}

\section{Introduction}

Large Language Models (LLMs) have advanced rapidly in recent years, with their applications in the field of software engineering becoming increasingly widespread \citep{DBLP:conf/acl/ZanCZLWGWL23,DBLP:journals/corr/abs-2311-10372,DBLP:journals/corr/abs-2311-07989,chen2024deep}.
Currently, LLMs' applications in software engineering have gradually expanded tasks at the code line and function level to more challenging repository-level tasks \citep{DBLP:conf/emnlp/ZhangCZKLZMLC23,DBLP:journals/corr/abs-2406-07003}. Within repository-level tasks, GitHub issue resolving is a key challenge, where LLMs are tasked to resolve the issue based on the issue description and the defective codebase. \citep{DBLP:conf/iclr/JimenezYWYPPN24,DBLP:journals/corr/abs-2407-01489}. This task can accelerate program repair and is crucial for improving development efficiency.

\begin{figure}[t]
  \includegraphics[width=\columnwidth]{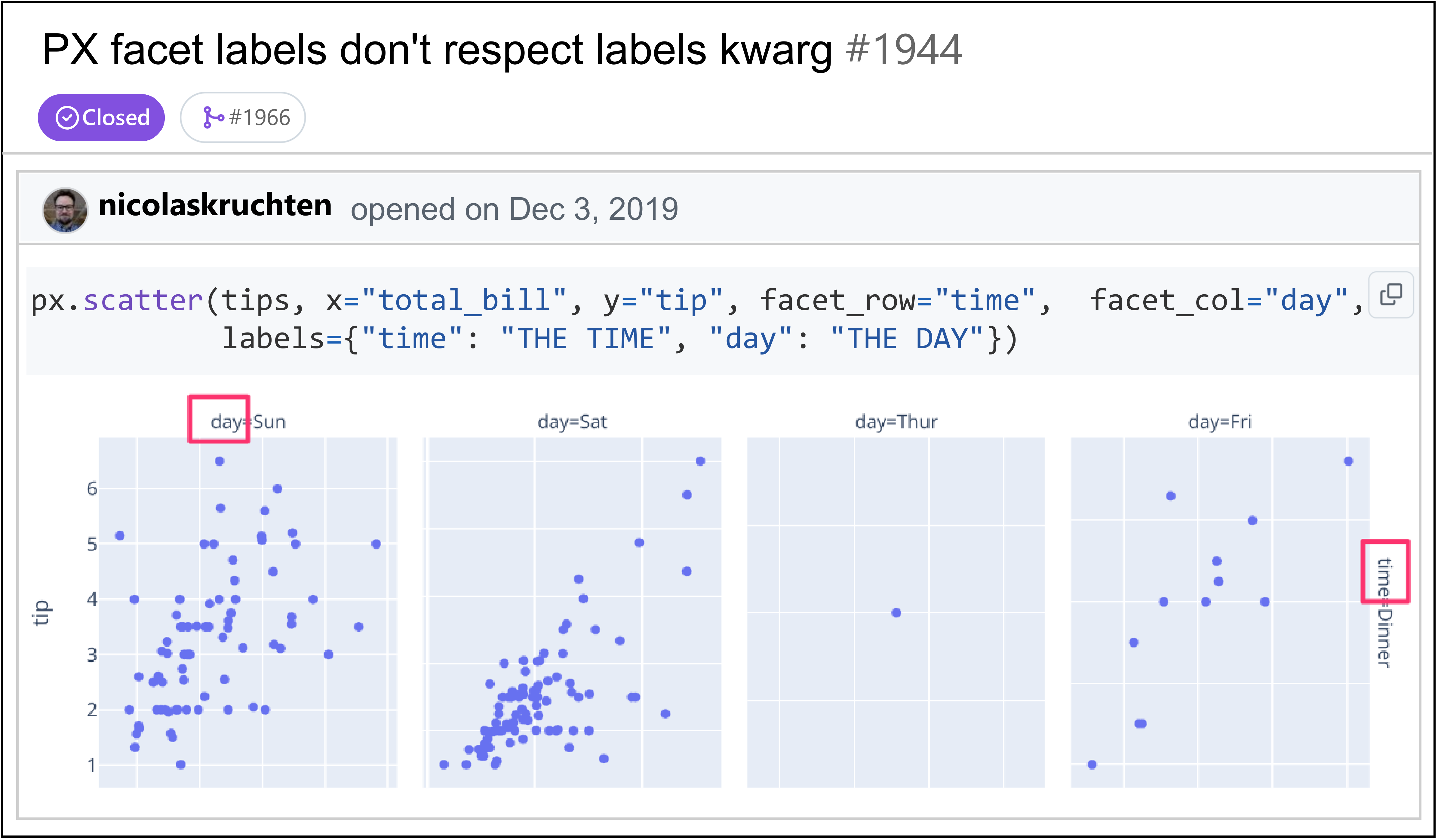}
  \caption{An example of a visual GitHub issue from \href{https://github.com/plotly/plotly.py/issues/1944}{Plotly issue \#1944}. The visual data illustrates that the label parameters (“time” and “day”) do not take effect.}
  \label{fig:importissue}
\end{figure}

Although recent approaches have made progress on this task, they focus exclusively on textual data in GitHub issues, neglecting visual data such as screenshots, diagrams, and videos \citep{DBLP:journals/corr/abs-2406-01304,DBLP:journals/corr/abs-2405-15793,DBLP:conf/issta/0002RFR24,DBLP:journals/corr/abs-2407-01489}.
However, this visual data is crucial for resolving issues, as it conveys additional knowledge that text alone cannot, including actual results, expected results, and error messages. Figure~\ref{fig:importissue} shows a specific example where visual data illustrates the running result. 
Moreover, we statistically analyze SWE-bench \citep{DBLP:conf/iclr/JimenezYWYPPN24}, the most popular benchmark for issue resolving. 
The result shows that over 5\% of GitHub issues contain visual data, with even higher percentages in visualization libraries like seaborn\footnote{\url{https://github.com/mwaskom/seaborn}} and matplotlib\footnote{\url{https://github.com/matplotlib/matplotlib}}, where they reach 45.5\% and 27.2\% respectively.
This analysis further highlights the importance of resolving visual GitHub issues. 
However, existing approaches struggle to resolve them effectively, as they overlook visual data, which calls for new solutions that leverage visual data.

An intuitive approach is to extract visual data from the issue and include it as part of the prompt.
While this approach seems to leverage visual data, it requires models with advanced multimodal and coding capabilities. Currently, only the latest commercial models, GPT-4o \citep{gpt4o} and Claude 3.5 Sonnet \citep{anthropic2024claude35}, barely meet these requirements, but their capabilities remain highly limited. Moreover, these models are less suitable for issue resolving due to high computational costs. 
Based on our analysis, using these models within the popular SWE-agent approach \citep{DBLP:journals/corr/abs-2405-15793} to run through all issues in SWE-bench once is estimated to cost an average of over \$4,700 \citep{DBLP:journals/corr/abs-2407-01489}, which imposes a significant financial burden on researchers.
To address this, we propose \codev, the first approach that leverages visual data to enhance the issue-resolving capabilities of LLMs at low cost. To resolve each issue, \codev follows a two-phase process: data processing and patch generation. In the data processing phase, \codev processes the visual data within the issue from both local and holistic perspectives. This phase produces fine-grained descriptions of the visual data and a structured summary of the entire issue. In the patch generation phase, \codev leverages the processed information to assist LLMs in generating a patch to resolve the issue.

To evaluate our approach, we construct a benchmark specifically designed for evaluating visual GitHub issue resolving, called Visual SWE-bench. The benchmark comprises 133 task instances spanning 11 open-source GitHub repositories, each of which has undergone rigorous selection. Finally, we conduct a series of experiments to validate the effectiveness of our approach. Experimental results demonstrate that \codev achieves a round 63.13\% relative improvement in the percentage of resolved instances on Visual SWE-bench compared to Agentless. Additionally, through case studies, we analyze the role of each component of \codev, providing insights into leveraging visual data to resolve issues. Overall, the contributions of this paper are as follows:

\begin{itemize}
\item We propose \codev, a simple yet novel approach that leverages visual data to enhance the issue-resolving capabilities of LLMs. 

\item We construct a benchmark designed to evaluate the performance of LLMs in resolving visual GitHub issues, namely Visual SWE-bench. The benchmark comprises 133 realistic software engineering tasks sourced from 11 open-source GitHub repositories.

\item We validate the effectiveness of our approach through a series of experiments and conduct in-depth analysis and summarization of the experimental results.
\end{itemize}




\section{Approach}
\begin{figure*}[t]
  \includegraphics[width=\linewidth]{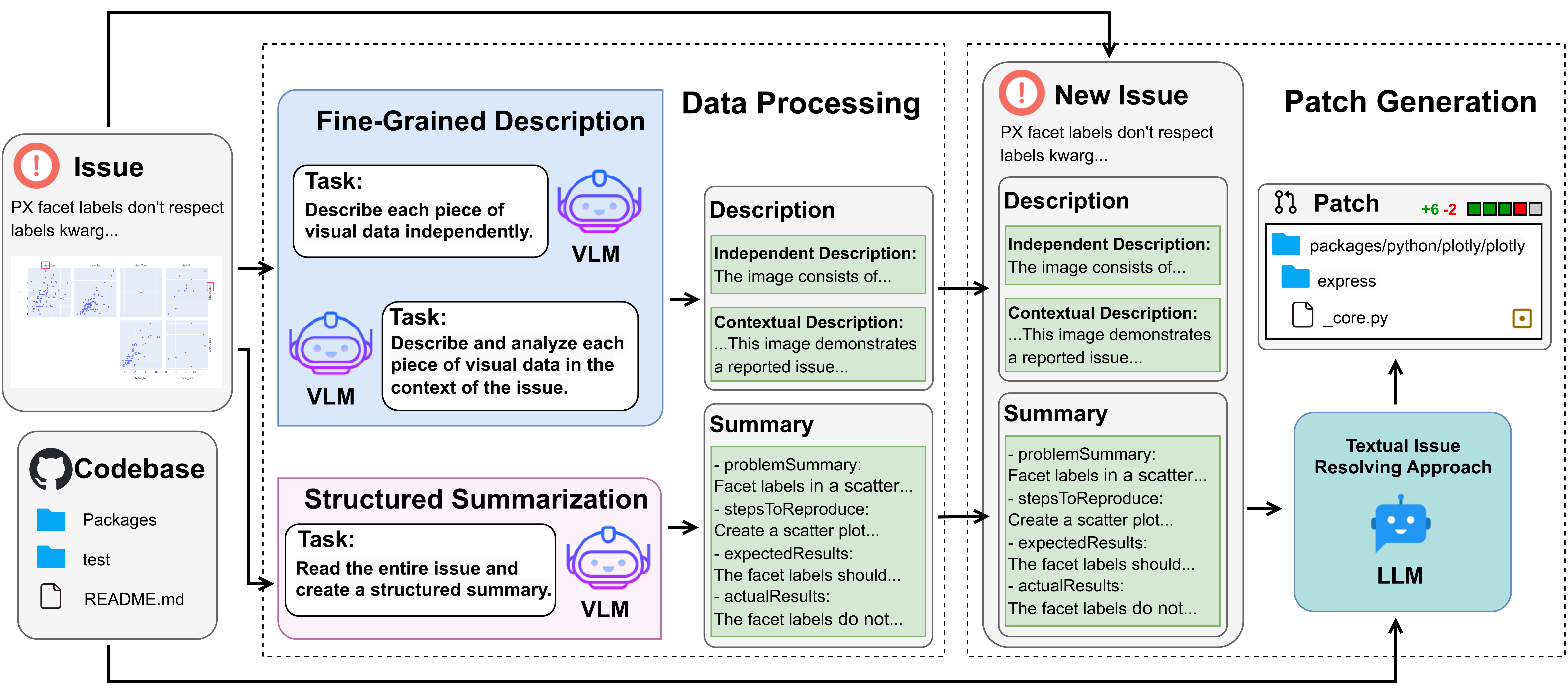}
  \caption{Overview of \codev.}
  \label{fig:overview}
\end{figure*}

Figure~\ref{fig:overview} illustrates an overview of \codev, which consists of two phases: data processing and patch generation. 
The first phase processes visual data and the second phase uses the processed information to assist LLMs in generating a patch. 
Below is a detailed description of each phase.
\subsection{Data Processing}
To process the issue's visual data, we adopt two components: fine-grained description and structured summarization. For fine-grained description, the Vision-Language Model (VLM) first generates an independent description for each piece of visual data based on its content. It then provides a contextual description that relates this data to the issue, resulting in a fine-grained description. 
In structured summarization, the VLM produces a summary that breaks down the complex issue into several clear sections. Below, we detail the implementation of each component.

\subsubsection{Fine-Grained Description}
We design the fine-grained description component to generate textual representations of the visual data. This component draws inspiration from how humans process visual data. When encountering visual data in an issue, humans first identify its raw features and then analyze its function in the context of the problem. Inspired by this, we design a two-step process to generate fine-grained descriptions.

\textbf{Step 1: Independent Description.}
In the first step, we instruct the VLM to describe each piece of visual data based solely on its content. For visual data consisting purely of text, such as a screenshot of error logs, the VLM extracts the text and outputs it in Markdown format. For other types of visual data, like images or videos with non-textual content, the VLM generates a detailed description capturing all details.

\textbf{Step 2: Contextual Description.}
In the second step, the VLM is prompted with the complete problem statement to establish the context. Subsequently, it is tasked with providing a comprehensive description and analysis of the visual data based on this contextual understanding. During this step, the VLM tightly connects the visual data with the problem’s context to analyze its function.

Through these two steps, we obtain fine-grained descriptions of all visual data in the issue. These descriptions capture not only the intrinsic features of the visual data but also its critical function in the problem’s context.

\subsubsection{Structured Summarization}
Some GitHub issues are described in a structured format, including reproduction steps, expected results, actual results, and so on. This format enhances clarity and makes the key aspects of issues easier to understand. Inspired by this, we propose generating a structured summary to enrich issues and reduce the difficulty of understanding them.

In the structured summarization component, the VLM is prompted with the complete problem statement, including the visual data. It is then tasked with understanding and analyzing the issue to generate a structured summary. To guide this process, we supply the VLM with a template that consists of the following fields: a brief problem summary, background information, reproduction steps, expected results, actual results, descriptions of visual data, and additional notes. However, not all issues may fit perfectly with this template. Therefore, we allow the VLM to skip irrelevant or unclear fields. Additionally, the summary can also include new fields if needed, as long as it remains clear and useful for resolving the issue.

Unlike fine-grained description, which focuses on generating representations of visual data, structured summarization aims to provide an overview of the entire issue. It not only covers visual data but also gives a deeper understanding of the problem. These two components complement each other: fine-grained description captures the detailed features of local visual data, while structured summarization synthesizes global information. Through these components, we ensure that visual data is processed effectively to support LLMs in understanding and resolving the issue. Prompts related to these components are listed in Appendix \ref{app:A}.

\subsection{Patch Generation}
After generating fine-grained descriptions and a structured summary in the data processing phase, the patch generation phase leverages this information to generate a patch. To support LLMs in efficiently utilizing this information, we splice them into the original issue. Specifically, the visual data is converted into fine-grained descriptions, and the issue is enriched with a structured summary, with an example provided in Appendix \ref{app:B}.

To enhance the ability of LLMs to resolve textual issues, various approaches have been proposed. These approaches take different forms: some are \textit{agent-based}, equipping LLMs with a set of tools that allow the agent to autonomously perform actions \citep{DBLP:journals/corr/abs-2406-01304,DBLP:journals/corr/abs-2405-15793,DBLP:conf/issta/0002RFR24}; others are \textit{agentless} \citep{DBLP:journals/corr/abs-2407-01489}. Regardless of their form, they typically input the issue and codebase, with the output being a generated patch. \codev combines these approaches through a unified interface, automating patch generation. At this point, the newly generated issue and its corresponding codebase are fed into the textual issue-resolving approach, where LLMs follow predefined instructions to generate a patch.

\section{Visual SWE-bench Benchmark}
In current benchmarks for the issue-resolving task, only the recently released SWE-bench Multimodal \citep{DBLP:journals/corr/abs-2410-03859} focuses on visual issues. However, as of writing, SWE-bench Multimodal\footnote{\url{https://www.swebench.com/multimodal.html}} lacks evaluation fields, and its evaluation script has not been made public, making it unsuitable for evaluation. To evaluate CodeV, we construct a benchmark for resolving visual GitHub issues, namely Visual SWE-bench. Below, we detail our benchmark construction process and its key features.
\subsection{Construction}
From the 2,294 instances in SWE-bench, we identify 128 task instances whose problem statement contains visual data. These visual data are presented through hyperlinks, with images embedded using HTML or Markdown syntax and videos provided as plain text hyperlinks. Building on these identified task instances, we adopt a four-stage construction process to further expand the tasks and conduct rigorous verification on all instances.

\begin{enumerate}

\item \textbf{Repositories selection and pull requests Collection.} 
We analyze the 128 task instances from SWE-bench, and the results show that most of them originate from visualization libraries. To expand our benchmark, we select three additional popular open-source visualization libraries (\href{https://github.com/plotly/plotly.py}{plotly.py}, \href{https://github.com/networkx/networkx}{networkx}, and \href{https://github.com/vega/altair}{altair}) and crawl all their pull requests (PRs) from GitHub. 
Since SWE-bench only includes PRs created before August 2023, we select repositories with at least 10 visual task instances from SWE-bench (\href{https://github.com/matplotlib/matplotlib}{matplotlib}, \href{https://github.com/sympy/sympy}{sympy}, \href{https://github.com/sphinx-doc/sphinx}{sphinx}, and \href{https://github.com/mwaskom/seaborn}{seaborn}) and collect recent PRs from these repositories. These two rounds of collection yield approximately 10,000 PRs.

\item \textbf{Candidate instance construction.} 
Candidate instances are constructed from the collected PRs through the following steps:
\begin{itemize}
\item[(1)] We select only merged PRs that resolve at least one issue and include modifications to test files.
\item[(2)] For each PR, we extract the text of all resolved issues, retaining only those PRs where the issue text contained hyperlinks to images or videos.
\item[(3)] For qualifying PRs, we gather detailed information, including “instance ID”, “patch”, “test patch”, and so on.
\end{itemize}
This process results in 38 candidate instances from approximately 10,000 PRs.

\item \textbf{Execution verification.} 
For each candidate instance, we meticulously set up the runtime environment and testing commands, removing any instances that failed due to installation or runtime errors. Next, we apply the test patch to each instance and record the test results both before and after applying the gold patch. Instances without any tests where the status changes from fail to pass are excluded. This process leaves 31 viable candidate instances.

\item \textbf{Human verification.} 
We conduct human verification on 159 instances, comprising 128 task instances from SWE-bench and 31 candidate instances filtered through the previous stages. Each instance is evaluated based on the following criteria: 
\begin{itemize} 
\item[(1)] Whether the visual data can be fully converted to text.
\item[(2)] Whether the visual data is essential for resolving the instance.
\item[(3)] Whether the problem description contains sufficient information for effective resolution. 
\end{itemize} 
Using these criteria, we exclude 4 instances where visual data can be fully converted to text via Optical Character Recognition (OCR) and 4 instances where visual data is not essential for resolution. Additionally, based on the content of their test cases, we exclude 18 instances that cannot be resolved due to insufficient problem descriptions. This process results in a curated, high-quality benchmark of 133 task instances.
\end{enumerate}
\subsection{Features}
\begin{figure}[t]
  \includegraphics[width=\columnwidth]{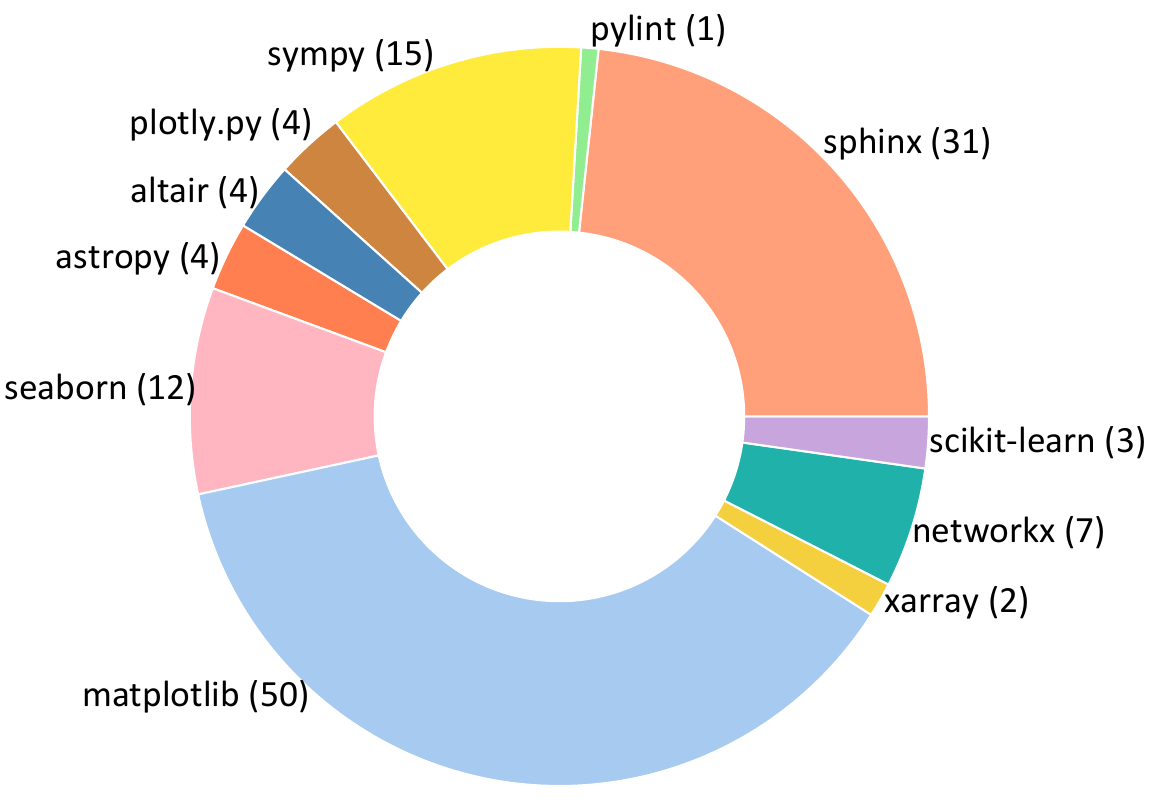}
  \caption{Distribution of Visual SWE-bench task instances across 11 open-source GitHub repositories.}
  \label{fig:SWE-bench-M-p}
\end{figure}

As shown in Figure~\ref{fig:SWE-bench-M-p}, Visual SWE-bench comprises 133 visual task instances sourced from 11 open-source GitHub repositories. These instances cover a wide range of functionalities, including but not limited to data visualization, machine learning, and document generation. This diverse set of tasks provides a comprehensive benchmark for evaluating the performance of LLMs in resolving visual issues automatically.
\begin{table*}
  \centering
  \resizebox{\textwidth}{!}{
  \begin{tabular}{lccccccccc}
\hline
  \multirow{2}{*}{\textbf{Repository}} & \multicolumn{2}{c}{\textbf{CodeBase}} 
  &  \multicolumn{1}{c}{\textbf{Issue Text}} 
  & \multicolumn{3}{c}{\textbf{Gold Patch}} 
  &  \multicolumn{1}{c}{\textbf{Tests}} 
  & \multicolumn{2}{c}{\textbf{Images}} \\
  \cmidrule(r){2-3}
  \cmidrule(r){4-4}
  \cmidrule(r){5-7}
  \cmidrule(r){8-8}
  \cmidrule(r){9-10}
    & \multicolumn{1}{c}{\textbf{\# Files}}
    & \multicolumn{1}{c}{\textbf{\# Lines}}
    & \multicolumn{1}{c}{\textbf{\# Length}}	
    & \multicolumn{1}{c}{\textbf{\# Lines}}	
    & \textbf{\# Files}	
    & \textbf{\# Func.}	
    & \multicolumn{1}{c}{\textbf{\# Lines}} 
    & \textbf{\# File Size} 
    & \textbf{\# Resolution}  \\
  \hline
\href{https://github.com/vega/altair}{altair}	&499	&90K	&98.5	&27	&2.5	&3	&15 &24.12	&99K \\
\href{https://github.com/astropy/astropy}{astropy}	&1578	&445K	&1352.5	&10	&1	&2	&69	&19.76	&170K\\
\href{https://github.com/matplotlib/matplotlib}{matplotlib}	&2388	&592K	&175	&9	&1	&2	&89.5	&23.58	&307K\\
\href{https://github.com/networkx/networkx}{networkx}	&849	&108K	&155	&23	&1	&1	&26	&21.38	&307K\\
\href{https://github.com/plotly/plotly.py}{plotly.py}	&14302	&587K	&44	&15	&2	&1	&11.5	&23.58	&307K\\
\href{https://github.com/pylint-dev/pylint}{pylint}	&2712	&92K	&100	&126	&4	&9	&10	&23.46	&307K\\
\href{https://github.com/scikit-learn/scikit-learn}{scikit-learn}	&1343	&277K	&641	&22	&1	&2	&12	&23.58	&307K\\
\href{https://github.com/mwaskom/seaborn}{seaborn}	&295	&70K	&143.5	&13.5	&2	&3	&109.5	&23.1	&307K\\
\href{https://github.com/sphinx-doc/sphinx}{sphinx}	&1436	&308K	&157	&8	&1	&2	&35	&27.08	&286K\\
\href{https://github.com/sympy/sympy}{sympy}	&1907	&477K	&125	&21	&1	&2	&36	&24.26	&275K\\
\href{https://github.com/pydata/xarray}{xarray}	&320	&123K	&220.5	&5	&1	&2	&229.5	&25.08	&275K\\
  \hline
mean	&2512	&288K	&292	&25.40	&1.59	&2.63	&58.45	&23.54	&268K\\
max	&14302	&592K	&1352.5	&126	&4	&9	&229.5	&27.08	&307K\\
  \hline
  \end{tabular}
  }
  \caption{\label{tab:statistics}
Summary statistics for Visual SWE-bench. The term “CodeBase \# Files and \# Lines” denotes the total count of files and lines within the codebase. “Issue Text \# Length” indicates the median word count in the problem statement. “Gold Patch \# Lines, \# Files, and \# Func.” reflects the median number of lines, files, and functions modified per patch stored in the repository. “Tests \# Lines” signifies the median line count of code present in test cases. “Images \# File Size and \# Resolution” represents both the median image file size (KB) and resolution (pixels).
  }
\end{table*}
\paragraph{Data statistics.} 
Table~\ref{tab:statistics} summarizes key statistics for the repositories in Visual SWE-bench, emphasizing their diversity and representativeness. Repository sizes range from 295 to 14,302 files and 70K to 592K lines of code, illustrating structural variation. Problem statements vary widely in length, with median word counts from 44 to 1,353, reflecting differences in task comprehension demands. Gold patches show diverse modification scopes, with median changes spanning 5 to 126 lines, indicating varying solution complexities. This broad spectrum of task characteristics provides a robust benchmark for evaluating LLMs' performance in resolving visual issues.

\paragraph{Visual data distribution.}
Across all Visual SWE-bench tasks, we identify 217 images and 2 videos, spanning a diverse range of visual processing challenges grouped into seven categories. These include code screenshots (21), error messages (8), and system information (2), which are linked to specific code library entities to facilitate error identification.  Other categories include data visualizations (140), documentation results (33), function formulas (13), and keyboard shortcuts (2), illustrating challenges such as generating complex statistics and utilizing code functions within specific libraries. Additionally, two instances feature GIFs (\href{https://github.com/matplotlib/matplotlib/issues/19633}{matplotlib\_matplotlib-19763}) and videos (\href{https://github.com/matplotlib/matplotlib/issues/25608}{matplotlib\_matplotlib-25631}), providing more dynamic and detailed depictions of these challenges.

\section{Experiments}
\begin{table*}[t]
\setlength\tabcolsep{4pt}
\centering
\begin{tabular}{lll}
    \hline
    Approach & Model &  Resolved (\%) \\
    \hline
    \hline
    \multicolumn{3}{c}{Evaluation results on 111 instances from Visual SWE-bench} \\
    \hline
    Honeycomb \citep{honeycomb} \faLock{} & NA  & 10.81 (12)\\
    Amazon Q Developer Agent \citep{amazonqdeveloper} \faLock{}&  NA  & 9.01 (10) \\
    Factory Code Droid \citep{factorydroid} \faLock{} &  NA  & 9.01 (10) \\
    AutoCodeRover \citep{DBLP:conf/issta/0002RFR24}& GPT 4o (2024-05-13)  & 10.81 (12)\\
    AppMap Navie \citep{appmapnavie} \faLock{} &  GPT 4o (2024-05-13)  & 9.01 (10)\\
    SWE-agent \citep{DBLP:journals/corr/abs-2405-15793} & Claude 3.5 Sonnet  & 6.31 (7)\\
     &GPT 4 (1106) & 8.11 (9)\\
     &GPT 4o (2024-05-13)& 1.80 (2)\\
    RAG \citep{DBLP:conf/iclr/JimenezYWYPPN24} & Claude 3 Opus  & 2.70 (3)\\
     &Claude 2& 0.90 (1)\\
    \codev + Agentless (Ours) &Qwen2-VL-72B + DeepSeek-V2.5 & 11.71 (13)\\
    &Qwen2-VL-2B + Qwen2.5-Coder-32B & \textbf{13.51 (15)}\\
    &Qwen2-VL-7B + Qwen2.5-Coder-32B & \textbf{13.51 (15)}\\
    &Qwen2-VL-72B + Qwen2.5-Coder-32B & 11.71 (13)\\
    \hline
    \multicolumn{3}{c}{Evaluation results on all instances from Visual SWE-bench} \\
    \hline
    Agentless \citep{DBLP:journals/corr/abs-2407-01489} &DeepSeek-V2.5 & 6.02 (8)\\
     &Qwen2.5-Coder-32B & 7.52 (10)\\
    Agentless Plus &Qwen2-VL-72B & 0.75 (1)\\
    \codev + Agentless (Ours) &Qwen2-VL-72B + DeepSeek-V2.5 & 9.77 (13)\\
    &Qwen2-VL-2B + Qwen2.5-Coder-32B & \textbf{12.78 (17)}\\
    &Qwen2-VL-7B + Qwen2.5-Coder-32B & \textbf{12.78 (17)}\\
    &Qwen2-VL-72B + Qwen2.5-Coder-32B & 11.28 (15)\\
    \hline
\end{tabular}
\caption{Results on Visual SWE-bench. The 111 instances are the overlapping instances between Visual SWE-bench and SWE-bench. \faLock{} indicates closed-source approaches.}
\label{tab:mainresults}
\end{table*}
\subsection{Experimental Setup}
\paragraph{Models.}To execute \codev for resolving visual issues, two model types are required: a VLM for processing visual data and an LLM for generating patches. To demonstrate the effectiveness of \codev, we specifically avoid commercial models and use open-source models in our experiments. For the VLM, we select Qwen2-VL \citep{Qwen2VL}, a model renowned for its robust visual understanding capabilities, using three versions: 2B, 7B, and 72B. For the LLM, we choose two models: DeepSeek-V2.5 \citep{deepseekv2} and Qwen2.5-Coder-32B \citep{hui2024qwen2}, both recognized for their powerful coding capabilities. 

\paragraph{Baselines.}In the patch generation phase, \codev combines textual issue-resolving approaches. We specifically adopt the open-source Agentless approach \citep{DBLP:journals/corr/abs-2407-01489}, which resolves issues through a simple localization and repair process. We also compare \codev with several textual issue-resolving approaches, including open-source and closed-source commercial products. These approaches have demonstrated strong performance on SWE-bench. To further contrast with VLM-based approaches, we design Agentless Plus, a modified version of Agentless that supports VLMs in processing visual data in issues to resolve them.

\paragraph{Metrics.}We use Resolved (\%) as our evaluation metric. The metric represents the percentage of Visual SWE-bench instances that have been successfully resolved. 
More details about the experiments can be found in Appendix \ref{app:C}.

\subsection{Evaluation}
\subsubsection{Main Results}
Table~\ref{tab:mainresults} presents the results of all approaches. The results show that \codev significantly enhances the issue-resolving capabilities of the LLM by leveraging visual data. Compared to all benchmarks, \codev achieves the best performance. When combined with Agentless, \codev achieves over a 50\% relative improvement, whether using DeepSeek-V2.5 or Qwen2.5-Coder-32B to resolve issues. 
The performance of CodeV highlights the value of leveraging visual data to help LLMs understand and resolve issues.

\begin{figure}[tbp]
    \subfigure[All approaches on 111 instances.]{ 
    \label{fig:ven:a} 
    \includegraphics[width=0.47\linewidth]{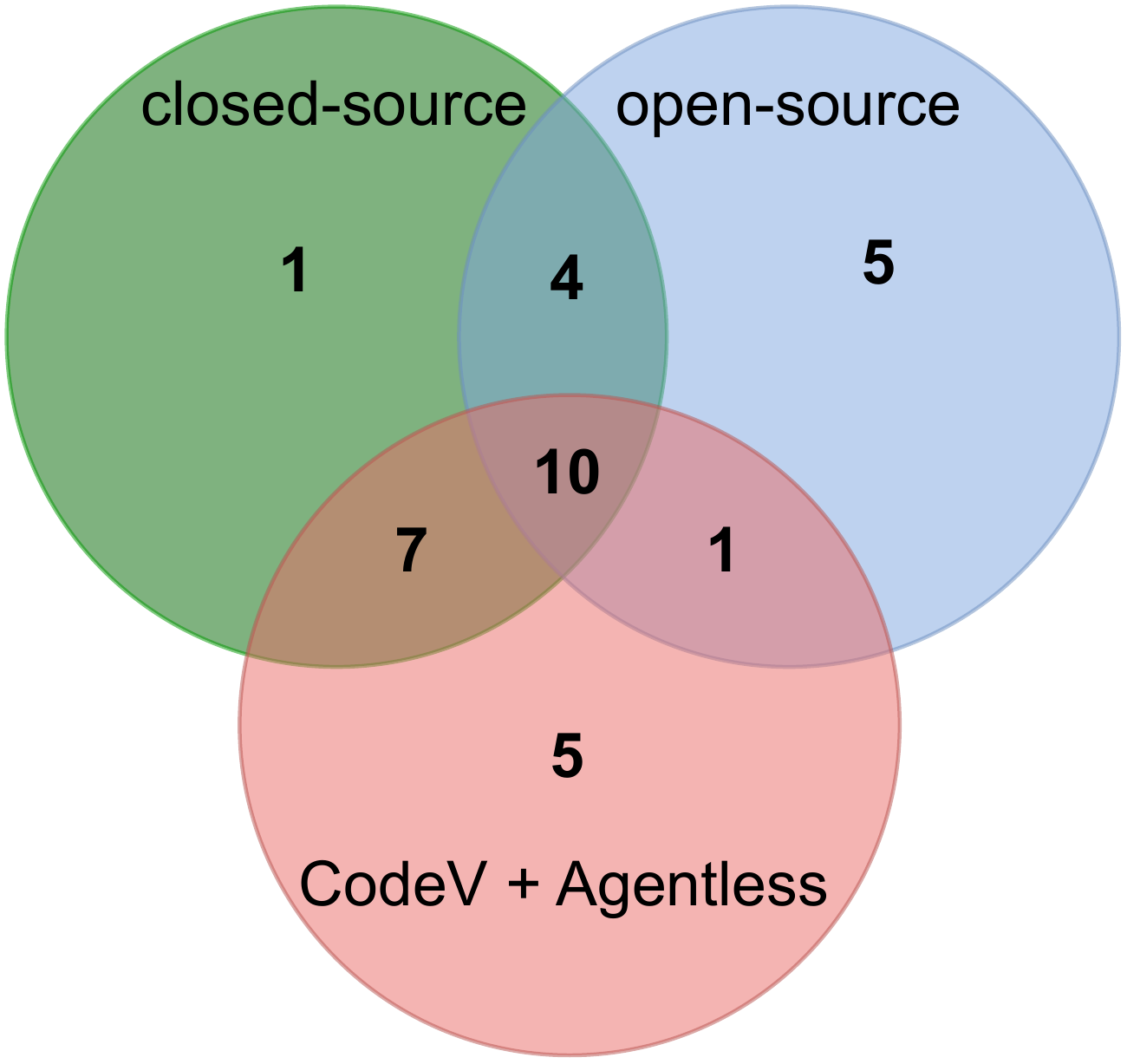}}\hfill
    \subfigure[\codev with VLMs of different model sizes on all instances.]{ 
    \label{fig:ven:b} 
    \includegraphics[width=0.47\linewidth]{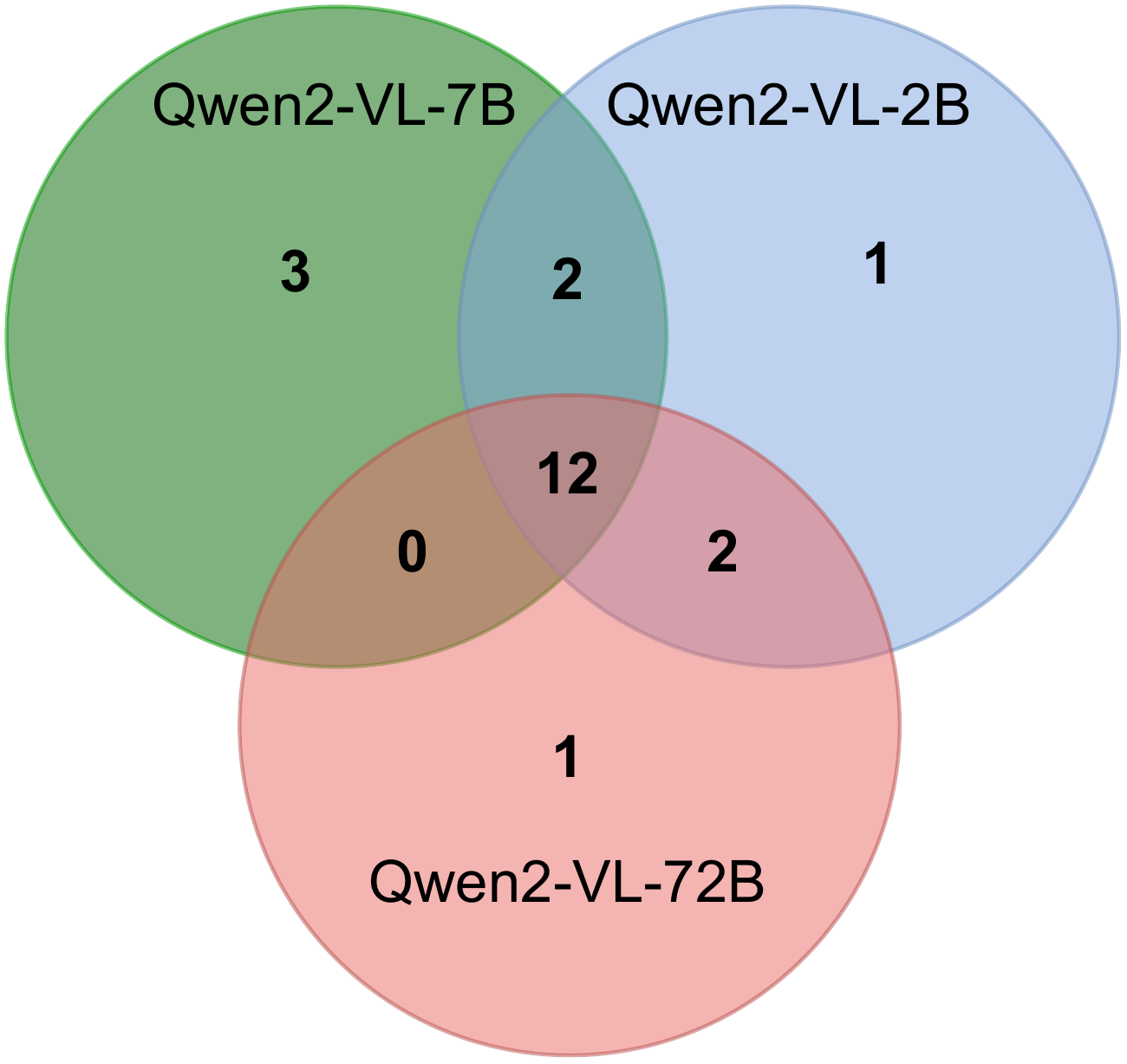}} 
  \caption{Venn diagrams of issues resolved from Visual SWE-bench.}
  \label{fig:ven}
\end{figure}

Figure~\ref{fig:ven:a} depicts the distribution of issues resolved by \codev compared to both closed-source and open-source baseline approaches. Notably, \codev can resolve certain issues that either open-source or closed-source approaches cannot resolve. Furthermore, \codev successfully resolves some complex issues that neither category of approaches could solve. This highlights not only the advantages of \codev but also the importance of leveraging visual data to resolve issues.

From Table~\ref{tab:mainresults}, it is evident that the performance of the VLM does not significantly impact \codev. Among the three versions of Qwen2-VL, the 72B model is the most powerful, while the 7B and 2B models exhibit progressively weaker capabilities. However, even with the lower-performing 7B and 2B models, \codev maintained robust issue-resolving capabilities, even outperforming the 72B model. Additionally, Figure~\ref{fig:ven:b} further illustrates the distribution of resolved issues across different VLMs. The issues resolved do not overlap entirely, indicating that each VLM has its strengths in processing different types of issues. This indicates that despite differences in VLM performance, \codev can still exert the capabilities of VLM, resolve issues stably, and demonstrate strong robustness.

Additionally, we observe that using the VLM alone, while it leverages visual data, does not yield satisfactory results. For example, Agentless Plus combined with Qwen2-VL-72B resolves only one issue. This is primarily due to its weak coding capabilities. In comparison, \codev effectively combines the VLM's visual understanding ability with the LLM's coding capabilities. This integration allows LLMs to leverage visual data to resolve issues at a low cost, making it a promising solution.
\subsubsection{Analysis of Ablation Studies}
\begin{table}
\setlength\tabcolsep{4pt}
\centering
\begin{tabular}{lll}
    \hline
    Approach &  Resolved (\%) \\
    \hline
\codev + Agentless & \textbf{11.28 (15)}\\
\hspace*{0.5em}w/o Fine-Grained Description   & 9.77 (13)\\
\hspace*{0.5em}w/o Independent Description   & 9.02 (12)\\
\hspace*{0.5em}w/o Contextual Description   & 9.02 (12)\\
\hspace*{0.5em}w/o Structured Summarization   & 7.52 (10)\\
    \hline
\end{tabular}
\caption{Ablation studies on Visual SWE-bench (133 instances). The VLM is Qwen2-VL-72B and the LLM is Qwen2.5-Coder-32B.}
\label{tab:ablationresults}
\end{table}
\begin{figure}[t]
  \includegraphics[width=\columnwidth]{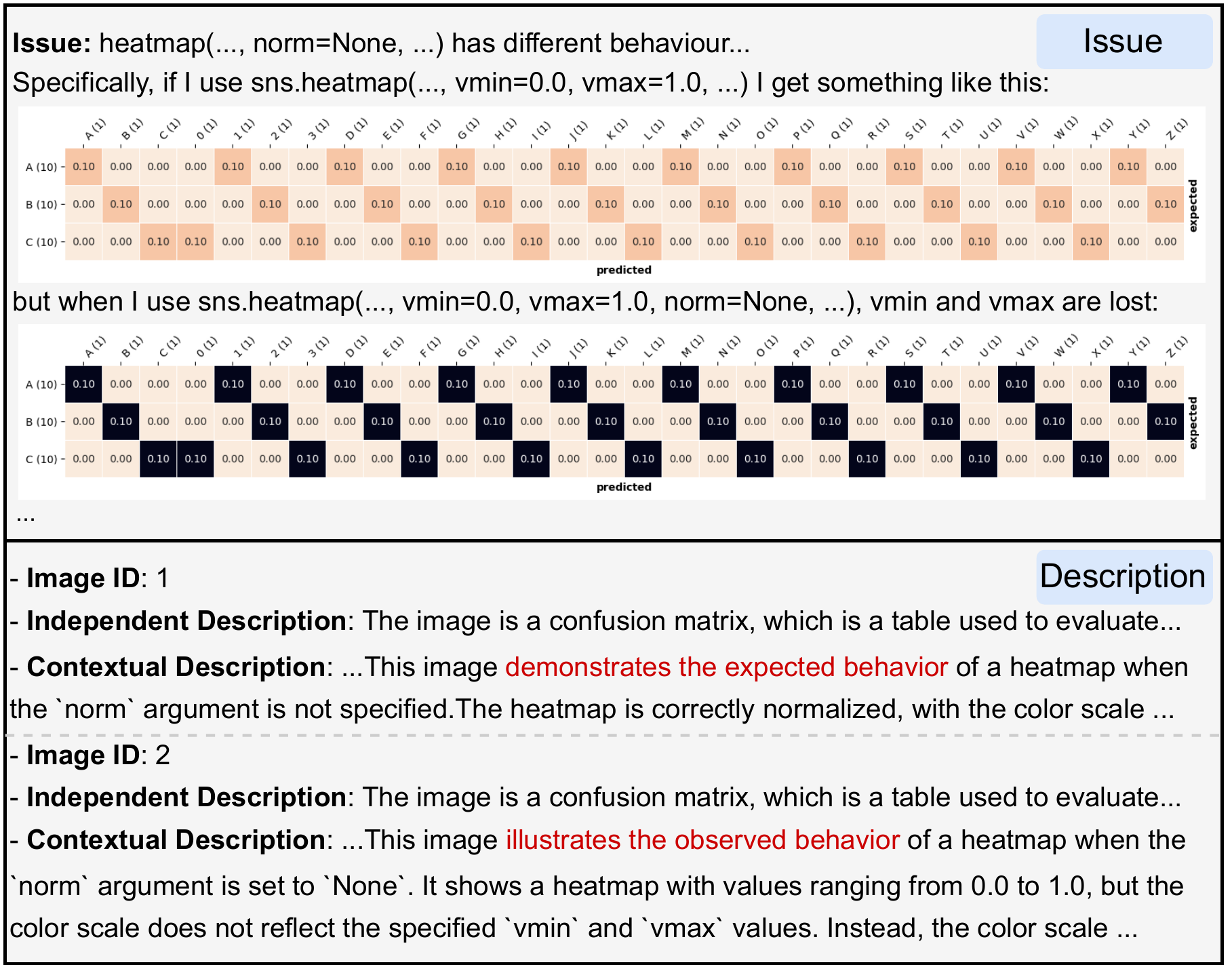}
  \caption{Fine-grained description example for the instance \href{https://github.com/mwaskom/seaborn/issues/3275}{mwaskom\_seaborn-3276}, offering detailed insights into the visual data.}
  \label{fig:F-G Description}
\end{figure}
We conduct a series of ablation studies on Visual SWE-bench, and the results in Table~\ref{tab:ablationresults} show that removing any component of \codev leads to a decline in performance. This led us to further investigate the functions of the components in the data processing phase.
\paragraph{Analysis of Fine-Grained Description.}
\begin{figure}[t]
  \includegraphics[width=\columnwidth]{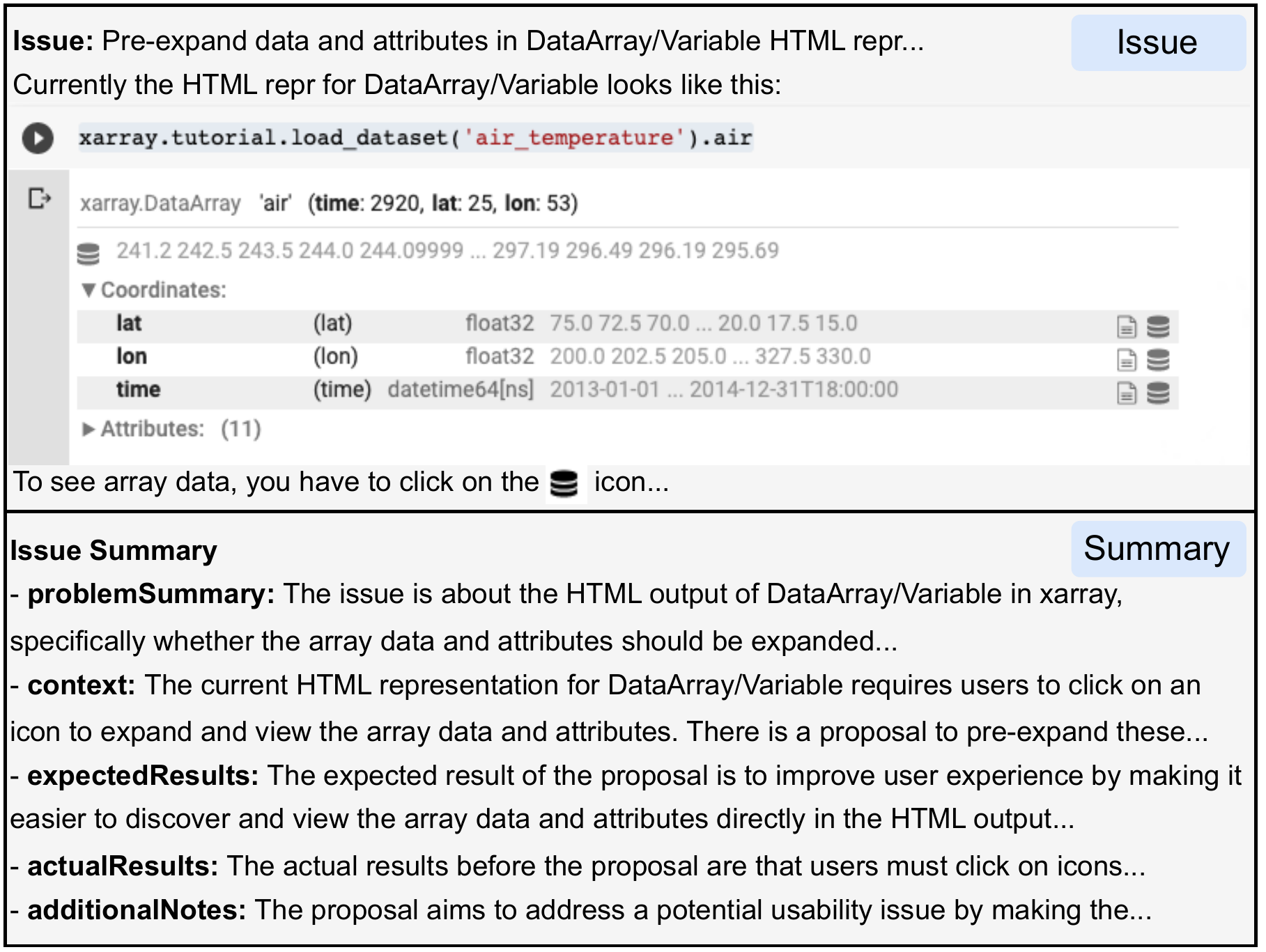}
  \caption{Structured summarization example for the instance \href{https://github.com/pydata/xarray/issues/4176}{pydata\_xarray-4182}, demonstrating a concise representation of its key information.}
  \label{fig:Summary}
\end{figure}
The fine-grained description process consists of two steps: independent description and contextual description. In the independent description, the VLM captures the raw features of visual data, providing a direct and detailed representation. However, why is contextual description also necessary? Figure~\ref{fig:F-G Description} shows an issue that is difficult to resolve without the contextual description. The figure shows two images, and the contextual description analyzes their respective function, explaining the information conveyed by each. In contrast, the independent description provides only a general overview, missing critical details needed for a complete understanding of the issue. These details are essential for LLMs to grasp the issue accurately. Thus, while the independent description captures the raw features of the visual data, the contextual description extracts deeper, more nuanced information. Together, these two steps work in tandem to provide a comprehensive understanding of the visual data.
\paragraph{Analysis of Structured Summarization.}
As shown in Table~\ref{tab:ablationresults}, removing structured summarization significantly undermines the performance of \codev. To explain this phenomenon, Figure~\ref{fig:Summary} presents an issue that is more easily resolved with a summary. 
The summary generated by \codev breaks down the complex issue into clear, digestible sections, providing LLMs with a full understanding of the issue’s background, expected outcomes, and actual results.
This structured format also helps LLMs grasp the core content more effectively. While the fine-grained description component attempts to convey the meaning of the visual data, relying solely on this still presents challenges in fully understanding the issue. By combining visual and textual data, the structured summary offers LLMs a more holistic understanding of the issue.


\section{Related Works}
\paragraph{Issue Resolving Approaches.} To assist LLMs in resolving GitHub issues, many approaches have already been proposed. Retrieval Augmented Generation (RAG) \citep{DBLP:conf/iclr/JimenezYWYPPN24} is a direct approach that resolves the issue by first extracting relevant code snippets from the repository and then using them to prompt LLMs to generate a patch. SWE-agent \citep{DBLP:journals/corr/abs-2405-15793} meticulously designs an agent-computer interface (ACI) that enables LLM agents to interact with repository environments to solve software engineering tasks. AutoCodeRover \citep{DBLP:conf/issta/0002RFR24} combines LLMs with code search, utilizes program structure, and conducts iterative searches for program improvement. CodeR \citep{DBLP:journals/corr/abs-2406-01304} is a multi-agent approach for issue-resolving tasks, adopting a multi-agent framework and pre-defined task graphs. Agentless \citep{DBLP:journals/corr/abs-2407-01489} points out the limitations of using agents and proposes a simple two-phase process of localization and repair to solve software development problems. However, these existing approaches overlook visual data within issues. \codev bridges this gap by processing visual data from both local and holistic perspectives, enhancing the capabilities of LLMs to resolve complex visual issues.

\paragraph{Code Generation Benchmarks.} Code generation has long been a measure of LLMs performance \citep{DBLP:journals/corr/abs-2108-07732}. The emergence of HumanEval \citep{DBLP:journals/corr/abs-2107-03374} provides a standardized framework for evaluating code generation models. In subsequent years, various benchmarks have been developed to enhance HumanEval by adding extensions to different languages \citep{cassano2022multipl,DBLP:conf/iclr/AthiwaratkunGWL23,DBLP:conf/icml/OrlanskiXGHHMAS23}, introducing variations in edit scope \citep{DBLP:conf/icse/YuSRZZMLLWX24,DBLP:journals/corr/abs-2308-01861}, presenting similar yet novel code completion tasks \citep{DBLP:conf/iclr/MuennighoffLZZH24}, and conducting more extensive testing \citep{DBLP:conf/nips/LiuXW023}. With the development of LLMs, existing benchmarks struggle to explore the boundaries of state-of-the-art LLMs' capabilities. To address this, SWE-bench \citep{DBLP:conf/iclr/JimenezYWYPPN24} offers a direction by researching real-world GitHub issues, serving as a challenging benchmark for evaluating next-generation LLMs. Building on this, SWE-bench-Java \citep{DBLP:journals/corr/abs-2408-14354} extends the benchmark to the Java ecosystem, creating a multilingual benchmark. Similarly, the latest work, SWE-bench Multimodal \citep{DBLP:journals/corr/abs-2410-03859} offers a multimodal upgrade to the benchmark, focusing on visual JavaScript problems. 
Given Python's increasing role in fields like data science, machine learning, and visualization, where visual data is crucial, we construct Visual SWE-bench focusing on visual issues in Python. 
By incorporating real-world visual issues, Visual SWE-bench encourages researchers to leverage visual data in solving complex software challenges.

\section{Conclusion}
We propose \codev, an approach that leverages visual data to resolve issues automatically. It processes visual data and provides LLMs with valuable information that enhances their ability to resolve issues. To evaluate \codev, we construct a benchmark for visual issue resolving, namely Visual SWE-bench. Through extensive experiments, we demonstrate the effectiveness of \codev and find that it maintains robust performance across VLMs with varying model sizes. Additionally, through case studies, we analyze the function of each component of \codev, offering insights on leveraging visual data to resolve issues.

\section*{Limitations}
Although this study offers valuable insights into leveraging visual data to resolve GitHub issues, several limitations should be acknowledged:
\begin{itemize}
\item Due to the randomness in the responses generated by LLMs, there is a potential threat to
the experimental results. Despite repeating each experiment twice to mitigate this, minor fluctuations in results may still occur.
\item Due to the lack of suitable benchmarks, our experiments are conducted solely on the self-constructed benchmark. 
However, we conduct comprehensive experiments and analyses to validate our approach, and we hope future research will develop more publicly available benchmarks to further explore this direction.
\item Due to the high costs of GPT-4o and Claude 3.5 Sonnet, we don't include them in our comparative experiments. Based on our estimates, using these models within the SWE-agent approach under similar experimental conditions would cost thousands of dollars. Nevertheless, we evaluate \codev using two LLMs and three VLMs, conducting extensive experiments that confirm its effectiveness.
\end{itemize}


\bibliography{ref}
\newpage
\appendix
\section*{Appendix}
\section{Prompts}
Figures~\ref{fig:prompt-1}–\ref{fig:prompt-4} show the prompts we use for images. The prompts for videos are almost identical, with "image" replaced by "video" in the text.
\label{app:A}
\subsection{Independent Description}
\begin{figure}[htbp]
  \includegraphics[width=\linewidth]{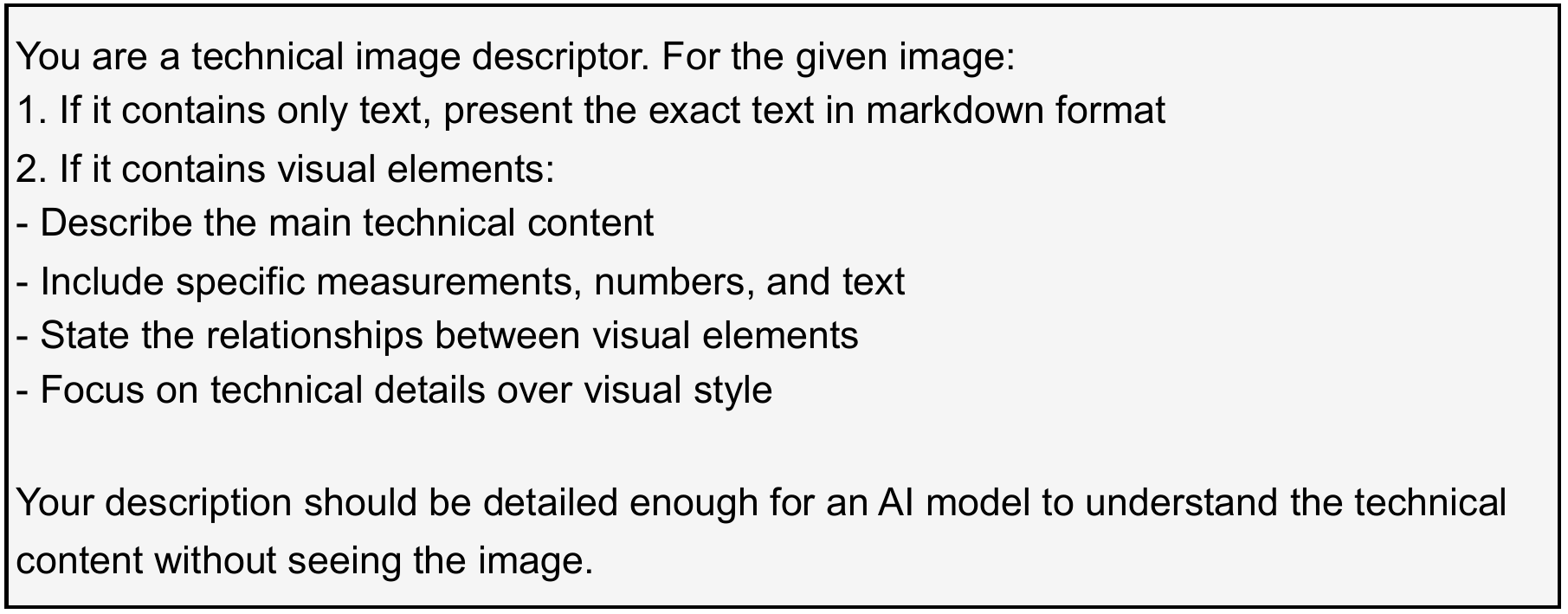}
  \caption{Prompt for generating independent descriptions of images.}
  \label{fig:prompt-1}
\end{figure}

Figure~\ref{fig:prompt-1} illustrates the prompt we use to generate independent descriptions, instructing the VLM to provide descriptions based on the image content.
\subsection{Contextual Description}
\begin{figure}[htbp]
  \includegraphics[width=\linewidth]{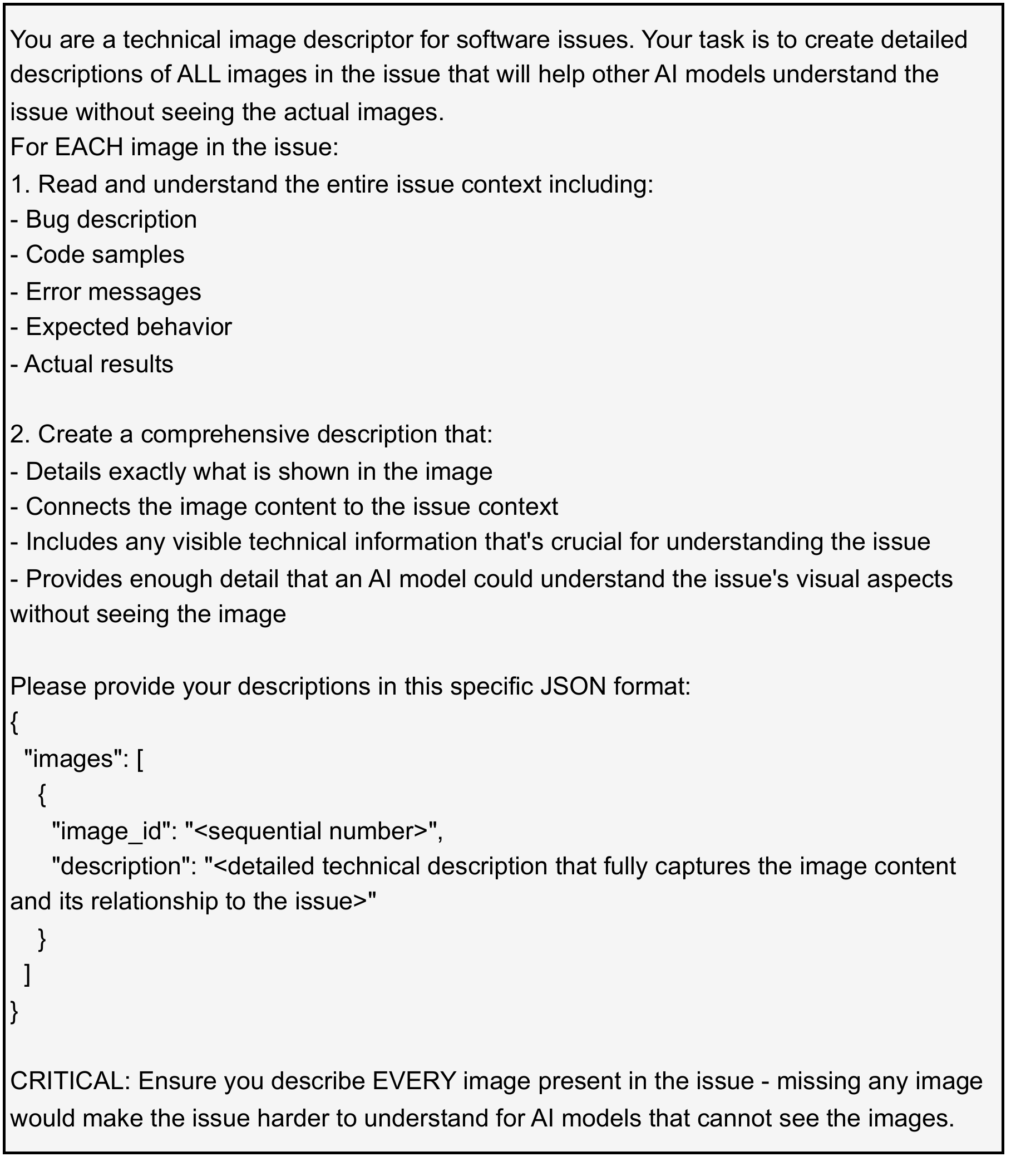}
  \caption{Prompt for generating descriptions of images based on the contextual information.}
  \label{fig:prompt-2}
\end{figure}

\begin{figure}[htbp]
  \includegraphics[width=\linewidth]{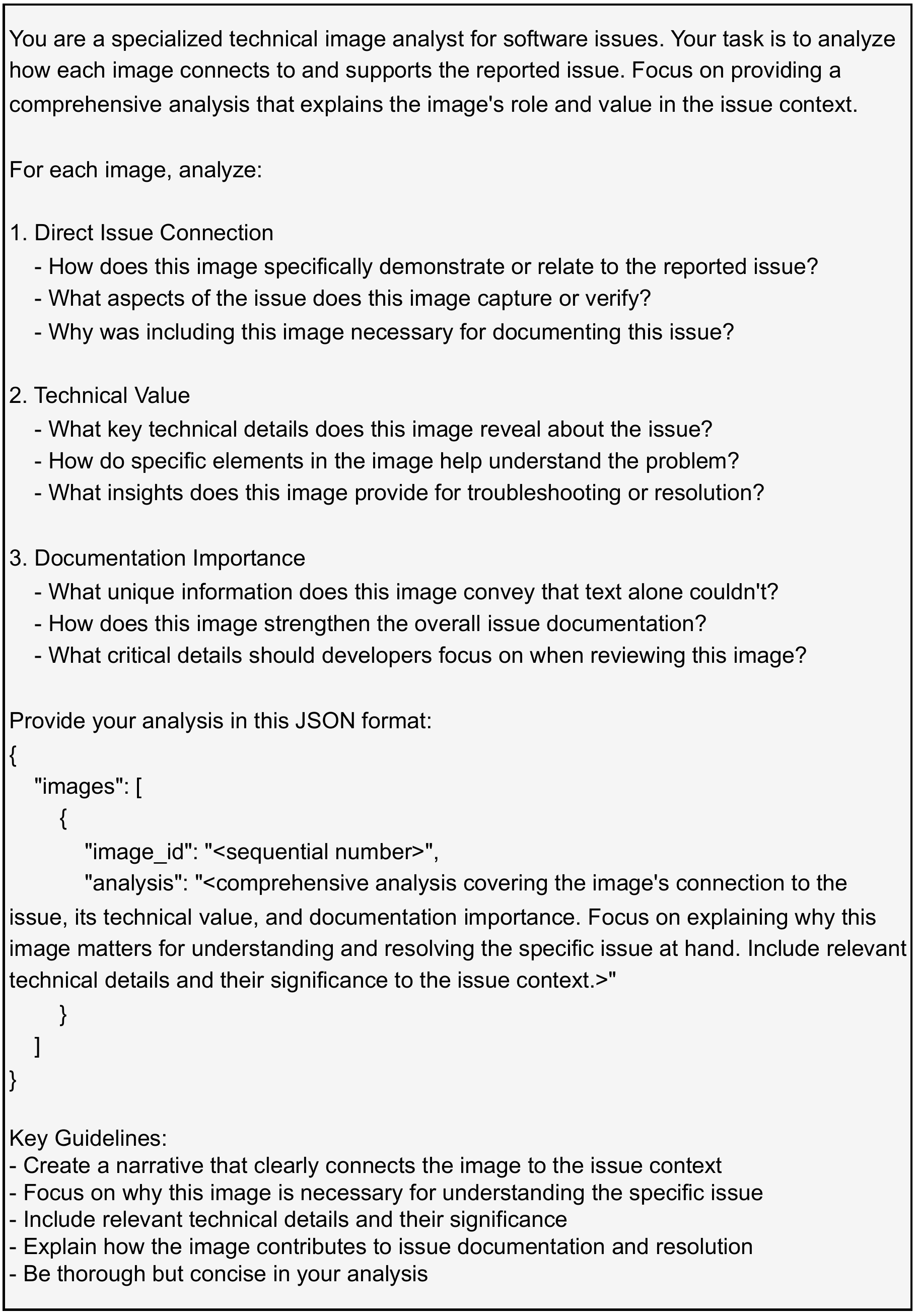}
  \caption{Prompt for analyzing the function of images within the given issue.}
  \label{fig:prompt-3}
\end{figure}

Figure~\ref{fig:prompt-2} and Figure~\ref{fig:prompt-3} present the prompts used for generating contextual descriptions. The prompt in Figure~\ref {fig:prompt-2} instructs the VLM to describe images based on the contextual information, while the prompt in Figure~\ref {fig:prompt-3} guides the VLM to analyze the function of images. 

\subsection{Structured Summary}
\begin{figure}[htbp]
  \includegraphics[width=\linewidth]{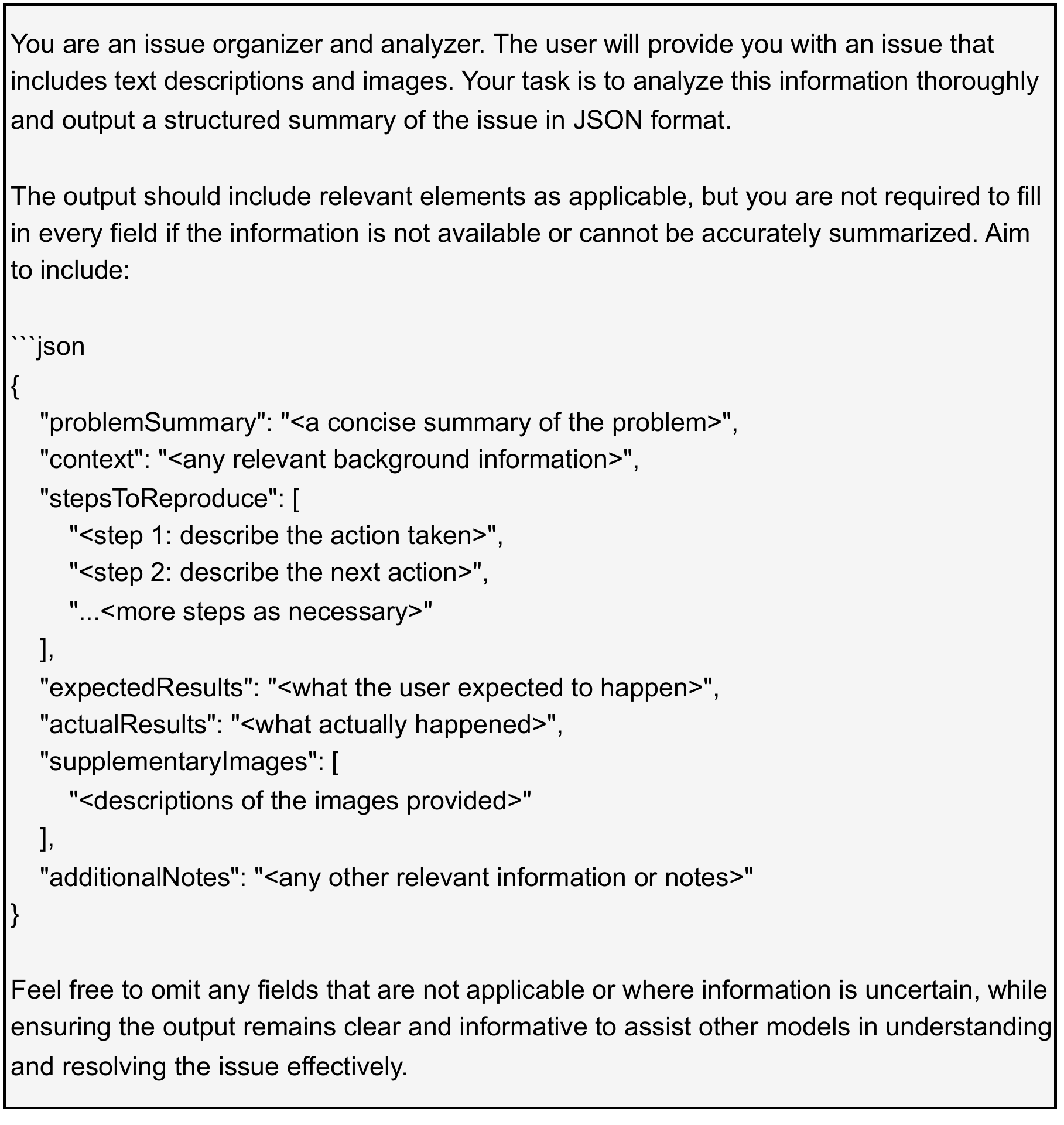}
  \caption{Prompt for generating a structured summary.}
  \label{fig:prompt-4}
\end{figure}

Figure~\ref {fig:prompt-4} illustrates the prompt designed for generating a structured summary. It instructs the VLM to produce a summary of the issue based on a referenced format.
\section{Example}
\label{app:B}
\begin{figure}[htbp]
  \includegraphics[width=\linewidth]{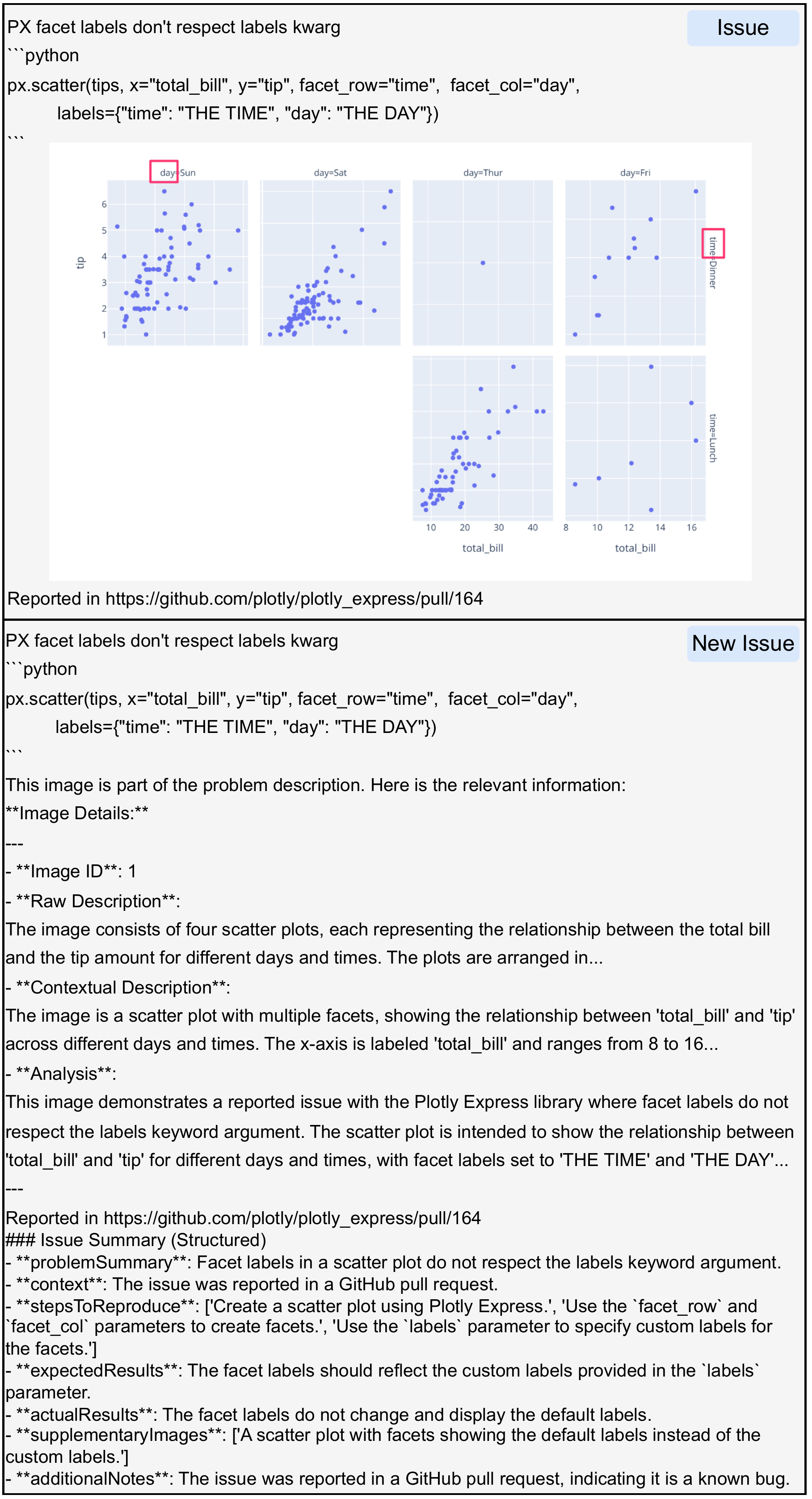}
  \caption{An example of a processed visual issue. The issue from \href{https://github.com/plotly/plotly.py/issues/1944}{Plotly issue \#1944}.}
  \label{fig:issue change}
\end{figure}

Figure~\ref {fig:issue change} presents an example where visual data is processed by the VLM, and the resulting information is appended to the original issue. 
\section{Other Experimental Details}
\label{app:C}
For models like Qwen2-VL and Qwen2.5-Coder-32B, we use vLLM for deployment on servers equipped with four NVIDIA H800 GPUs (each with 80GB of memory). For the DeepSeek-V2.5 model, we utilize the official API service provided by its developers. All experiments are conducted twice to determine the maximum number of instances that can be resolved. When using the Agentless approach, we employ version 1.0.
\end{document}